\documentclass[a4paper,12pt]{article}
\usepackage{graphicx}
\usepackage[small]{subfigure,epsfig}

\usepackage {amsmath} \usepackage{amssymb} \usepackage{cite}

\newcommand {\cosech} {cosech}

\DeclareMathOperator{\sech}{sech} 

\begin{document}

\author{Nikolai A. Kudryashov\footnote{E-mail: Kudryashov@mephi.ru} \, and  Dmitry I. Sinelshchikov}

\title{A note on "New abundant solutions for the Burgers equation"}

\date{Department of Applied Mathematics, National Research Nuclear University "MEPhI", Kashirskoe sh. 31, Moscow 115409, Russia}

\maketitle

\begin{abstract}

Salas, Gomez and Herana\'ndez [A.Y. Salas S., C.A. Gomez S., J.E.C Herna\'ndez, New abundant solutions for tha Burgers  equation, Computers and Mathematics with Applications 58 (2009) 514 -520] presented 70 "new exact solutions" of a "generalized version" of the Burgers equation. In this comment we show that all 70 solutions by these authors are not new and cannot be new.

\end{abstract}



\section{Introduction}

Recently Salas, Gomez and Herana\'ndez in \cite{Salas}  considered the Burgers equation in the form
\begin{equation}
u_{t}+\alpha\,u\,u_{x}+\beta\,u_{xx}=0.
\label{eq1}
\end{equation}

The authors \cite{Salas} believe that they studied a
"generalized version" of the Burgers equation but they are wrong here.
Eq.\eqref{eq1} can be transformed to the usual form of the Burgers equation \cite{Burgers, Rosenblatt68, Benton66}
\begin{equation}
u_{t}+u\,u_{x}=\beta\,u_{xx}
\label{eq3}
\end{equation}
if we use the following transformations
\begin{equation}
u=\frac{1}{\alpha}\,u',\quad x=-x',\quad t=-t'
\label{eq2}
\end{equation}
(primes in \eqref{eq3} are omitted).

Eq. \eqref{eq3} was firstly introduced in \cite{Bateman}. But this equation became popular after work  \cite{Burgers} for describing turbulence processes. It is well known that the Burgers equation can be linearized by the Cole---Hopf transformation \cite{Hopf51,Cole50}
\begin{equation}
u=-2\,\beta\,\frac{\partial \ln z}{\partial x}
\label{eq4}
\end{equation}
As a result of application of transformation \eqref{eq4}, we have
\begin{equation}
u_{t}+u\,u_{x}-\beta\,u_{xx}=-2\,\beta\frac{\partial}{\partial\,x}\left[\frac{z_{t}-
\beta\,z_{xx}}{z}\right]
\label{eq5}
\end{equation}
Thus, solution of the Burgers equation \eqref{eq3} can be expressed via solutions of the linear heat equation
 \begin{equation}
z_{t}-\beta\,z_{xx}=0
\label{heq}
\end{equation}
Solving the Cauchy problem for Eq. \eqref{heq} we can obtain the solution of the Cauchy problem for the Burgers equation \eqref{eq3} \cite{Whitham,Kudryashov}.

\section{Analysis of 70 exact solutions of the Riccati equation by Salas, Gomez and Herana\'ndez}

Taking a "new modified Exp-function method" into account Salas, Gomez and Herana\'ndez in \cite{Salas} have used the traveling wave $\xi=x+\lambda\,t$ for the Burgers equation \eqref{eq1} and obtained 70 solutions of the nonlinear ordinary differential equation
\begin{equation}
\lambda\,u'(\xi)+\alpha\,u(\xi)\,u'(\xi)+\beta\,u''(\xi)=0\,.
\label{eq6}
\end{equation}

At this stage the authors \cite{Salas} essentially reduced a class of possible solutions
for Eq.\eqref{eq1} because the authors studied the nonlinear ordinary differential equation \eqref{eq6} but not the partial differential equation \eqref{eq1}.

The authors \cite{Salas} did not note that Eq.\eqref{eq6} can be integrated. Integrating  Eq. \eqref{eq6} with respect to $\xi$ we obtain the famous Riccati equation
\begin{equation}
-C+\lambda\,u(\xi)+\frac{\alpha}{2}\,u(\xi)^{2}+\beta\,u'(\xi)=0,
\label{eq7}
\end{equation}
where $C$ is a constant of integration.

Equation \eqref{eq7} was introduced by Italian mathematician Jacopo Francesco Riccati in 1724. After that Eq. \eqref{eq7} was studied many times (see \cite{Glaisher, Reid, Davis, Kamke, Polyanin03, Polyanin07}).

The general solution of the Riccati equation is well known and is described by the
formulae ( see for example \cite{Kamke,Kudryashov} )

\begin{equation}
\begin{gathered}
u \left( \xi \right) = -\frac{\lambda}{\alpha}+ \frac{2\,\beta\,K}
{\alpha}\, \tanh \left\{K\,{
 \left( \xi+ C_{1} \right) } \right\},\quad K=\frac {
\sqrt {2\,C\alpha+{\lambda}^{2}} }{2\beta
},
 \\
 \\ \lambda^{2}+C^{2} \neq 0,
\label{eq8}
\end{gathered}
\end{equation}

\begin{equation}
\begin{gathered}
u \left( \xi \right)=\frac{2\beta}{\alpha\,\xi+2\beta\,C_{1}},\quad C=\lambda=0,
\label{eq9}
\end{gathered}
\end{equation}
where $C_1$ is an arbitrary constant.

These solutions were found more than one century ago and nobody can find new solutions of Eq. \eqref{eq7}.

An alternative form of expression \eqref{eq8} is
\begin{equation}
\begin{gathered}
u \left( \xi \right) = \frac{1}{\alpha}\,\left(2\,\beta\,K-\lambda-\frac{4\,\beta\,K}{1+C_2\,e^{2\,K\,\xi}}\right),
\label{eq8a}
\end{gathered}
\end{equation}
where $C_2=e^{2\,K\,C_1}$.

Solution \eqref{eq8a} follows from the set of identities
\begin{equation}
\begin{gathered}
u \left( \xi \right) =-\frac{\lambda}{\alpha}+ \frac{2\,\beta\,K}
{\alpha}\, \tanh \left\{K\,{
 \left( \xi+ C_{1} \right) } \right\}=\\
 \\
 =-\frac{\lambda}{\alpha}+ \frac{2\,\beta\,K}
{\alpha}\, \left(\frac{e^{K(\xi+C_1)}-e^{-K(\xi+C_1)}}{e^{K(\xi+C_1)}+e^{-K(\xi+C_1)}}\right)=\\
 \\
 =-\frac{\lambda}{\alpha}+ \frac{2\,\beta\,K}
{\alpha}\, \left(1-\frac{2\,e^{-K(\xi+C_1)}}{e^{K(\xi+C_1)}+e^{-K(\xi+C_1)}}\right)= \\
 \\
 =\frac{1}{\alpha}\,\left(2\,\beta\,K-\lambda-\frac{4\,\beta\,K}{1+C_2\,e^{2\,K\,\xi}}\right).
\label{eq8b}
\end{gathered}
\end{equation}

Following to the report by one of the referees let us show that all solutions by Salas, Gomez and Herana\'ndez in \cite{Salas} can be reduced to the formulae \eqref{eq8} or \eqref{eq8a}.

In \cite{Salas}, the solutions $u_{2m}$ $(m=1,\ldots,35)$ are obtained from the solutions $u_{2m-1}$ by replacing $\mu$ by $i\,\mu$. (There is typographical error in $u_8$: $'x+'$ should be $'x-'$.) Consequently, it is only necessary to show that solutions $u_{2m-1}$ $(m=1,\ldots,35)$ are just special cases of \eqref{eq8} or \eqref{eq8a}. We have

$u_1$ is  \eqref{eq8a} with $K=-{\mu}/{2}$,  $\lambda=-\beta\,\mu$, $C_2=b_2$;

$u_3$ is  \eqref{eq8a} with $K={\mu}/{2}$,  $\lambda=\beta\,\mu$, $C_2=b_2$;

$u_5$ is  \eqref{eq8a} with $K=-{\mu}/{2}$,  $\lambda=-(\beta\,\mu+p\,\alpha)$,  $C_2=b_2$;

$u_7$ is  \eqref{eq8a} with $K=-{\mu}/{2}$,  $\lambda=-(\beta\,\mu+\frac{a_2}{b_2}\,\alpha)$,  $C_2=b_2$;

$u_9$ is  \eqref{eq8a} with $K={\mu}/{2}$,  $\lambda=\beta\,\mu-\frac{a_2}{b_2}\,\alpha$,  $C_2=b_2$;

$u_{11}$ is  \eqref{eq8a} with $K=-{\mu}/{2}$,  $\lambda=-(\beta\,\mu \,+ p\,\alpha+\frac{a_2}{b_2}\,\alpha)$,  $C_2=b_2$;

$u_{13}$ is  \eqref{eq8a} with $K={\mu}/{2}$,  $\lambda=\beta\,\mu \,- p\,\alpha-\frac{a_2}{b_2}\,\alpha$,  $C_2=b_2$;

$u_{15}$ is  \eqref{eq8} with $K=\mu$,  $\lambda=- p\,\alpha$,  $K\,C_1=i\,\pi/2$;

$u_{17}$ is  \eqref{eq8} with $K=\mu$,  $\lambda=- p\,\alpha$,  $\,C_1=0$;

$u_{19}$ is  \eqref{eq8} with $K=\mu/2$,  $\lambda=- p\,\alpha$,  $K\,C_1=i\,\pi/2$;

$u_{21}$ is  \eqref{eq8} with $K=\mu/2$,  $\lambda=- p\,\alpha$,  $\,C_1=0$;

$u_{23}$ is  \eqref{eq8} with $K=\mu$,  $\lambda=- \frac{a_1}{b_1}\,\alpha$,  $K\,C_1=i\,\pi/2$;

$u_{25}$ is  \eqref{eq8} with $K=\mu$,  $\lambda=- \frac{a_2}{b_2}\,\alpha$,  $\,C_1=0$;

$u_{27}$ is  \eqref{eq8} with $K=\mu$,  $\lambda=- \left(p+\frac{a_1}{b_1}\right)\,\alpha$,  $K\,C_1=i\,\pi/2$;

$u_{29}$ is  \eqref{eq8} with $K=\mu$,  $\lambda=- \left(p+\frac{a_2}{b_2}\right)\,\alpha$,  $\,C_1=0$;

$u_{31}$ is  \eqref{eq8} with $K=\mu/2$,  $\lambda=- \frac{\beta\,\mu\,a_1}{a_2}\,$,  $\,C_1=0$;

$u_{33}$ is  \eqref{eq8} with $K=a_2\,\mu/2$,  $\lambda=- \frac{\beta\,\mu\,a_1}{a_2}\,$,  $K\,C_1=i\,\pi/2$;

$u_{35}$ is  \eqref{eq8} with $K=\mu/2$,   $2\,K\,C_1=\theta_0+i\,\pi$, where
 $\tanh{\theta_0}=\frac{p\,\alpha+\lambda}{\beta\,\mu}$;

$u_{37}$ is  \eqref{eq8} with $K=\mu/2$,  $2\,K\,C_1=\theta_0$, where
 $\tanh{\theta_0}=\frac{p\,\alpha+\lambda}{\beta\,\mu}$;

$u_{39}$ is  \eqref{eq8} with $K=\mu/2$,  $\lambda=-p\,\alpha$, $2\,K\,C_1=\theta_0+i\pi$, where
 $\tanh{\theta_0}=\frac{a_0\,\alpha}{\beta\,\mu}$;

$u_{41}$ is  \eqref{eq8} with $K=\mu/2$,  $\lambda=-p\,\alpha$, $2\,K\,C_1=\theta_0$, where
 $\tanh{\theta_0}=\frac{a_0\,\alpha}{\beta\,\mu}$;

$u_{43}$ is  \eqref{eq8a} with $K={\mu}/{2}$,  $\lambda=-(-\beta\,\mu \,+ p\,\alpha)$,  $C_2=b_2$;

$u_{45}$ is  \eqref{eq8} with $K=\mu/2$,  $\lambda=-p\,\alpha+i\beta\mu\,b_2$, $\,K\,C_1=\theta_0+i\pi/4$, where  $\tanh{\theta_0}=\frac{1}{i\,b_2}$;

$u_{47}$ is  \eqref{eq8} with $K=\mu/2$,  $\lambda=-p\,\alpha-i\beta\mu\,b_2$, $\,K\,C_1=-\theta_0 -i\pi/4$, where  $\tanh{\theta_0}=\frac{1}{i\,b_2}$;

$u_{49}$ is  \eqref{eq8} with $K=\mu/2$,   $2\,K\,C_1=\theta_0+i\pi$, where  $\tanh{\theta_0}=\frac{\lambda}{\beta\,\mu}$;

$u_{51}$ is  \eqref{eq8} with $K=\mu/2$,  $2\,K\,C_1=\theta_0$, where  $\tanh{\theta_0}=\frac{\lambda}{\beta\,\mu}$;

$u_{53}$ is  \eqref{eq8} with $K=\mu/2$,  $2\,K\,C_1=\theta_0$, where  $\tanh{\theta_0}=\frac{a_0\,\alpha+\lambda}{\beta\,\mu}$;

$u_{55}$ is  \eqref{eq8} with $K=\mu/2$,  $\lambda=i\,\beta\,\mu\,b_2$, $\,K\,C_1=\theta_0+i\,\pi/4$, where  $\tanh{\theta_0}=\frac{1}{i\,b_2}$;

$u_{57}$ is  \eqref{eq8} with $K=\mu/2$,  $\lambda=-i\,\beta\,\mu\,b_2$, $\,K\,C_1=-\theta_0-i\,\pi/4$, where  $\tanh{\theta_0}=\frac{1}{i\,b_2}$;

$u_{59}$ is  \eqref{eq8} with $K=\mu/2$,  $\lambda=- p\,\alpha$,  $K\,C_1=-i\,\pi/4$;

$u_{61}$ is  \eqref{eq8} with $K=\mu/2$,  $\lambda=- p\,\alpha$,  $K\,C_1=i\,\pi/4$;

$u_{63}$ is  \eqref{eq8} with $K=\mu/2$,  $\lambda=- p\,\alpha-\frac{\beta\,\mu\,a_2}{a_1}$,  $K\,C_1=-i\,\pi/4$;

$u_{65}$ is  \eqref{eq8} with $K=\mu/2$,  $\lambda=- p\,\alpha-\frac{\beta\,\mu\,a_2}{a_1}$,  $K\,C_1=i\,\pi/4$;

$u_{67}$ is  \eqref{eq8} with $K=\mu/2$,  $\lambda=-\frac{\beta\,\mu\,a_2}{a_1}$,  $K\,C_1=-i\,\pi/4$;

$u_{69}$ is  \eqref{eq8} with $K=\mu/2$,  $\lambda=-\frac{\beta\,\mu\,a_2}{a_1}$,  $K\,C_1=i\,\pi/4$.

In considering $u_{59}$, $u_{61}$, $u_{63}$, $u_{65}$, $u_{67}$ and $u_{69}$ it was used the identities
\begin{equation}
\tanh{z}-i\sech{z}=\coth{\left(z-\frac{i\pi}{2}\right)}-\cosech{\left(z-\frac{i\pi}{2}\right)}=
\tanh{\left(\frac {z}{2}-\frac{i\pi}{4}\right)}\end{equation}

Thus, the analysis of 'many new solutions' of the Riccati equation \eqref{eq6} shows that all 70 exact solutions by Salas, Gomez and Herana\'ndez \cite{Salas} can be found from the general solution \eqref{eq8} of Eq.\eqref{eq6}. At first glance we have the only negative moment of work \cite{Salas}. However the authors obtained 70 different forms of the solution of the Riccati equation. Taking the Riccati equation as the simplest equation in the method discussed in \cite{Kudryashov_2005a, Kudryashov_2005b} we can imagine how many methods can be suggested to search for exact solutions of nonlinear differential equations. Every form of the solution for the Riccati equation can be used in finding exact solutions of nonlinear ordinary differential equations. However we hope the researches will not use this dubious idea.

Salas, Gomez and Herana\'ndez wrote in \cite{Salas} "we conclude that the variant of the Exp - method here used is a very powerful mathematical tool for solving other nonlinear equations". However the analysis of the solutions for the Riccati equation by the paper \cite{Salas} points clearly to the obvious deficiency of the Exp - function method in finding exact solutions of nonlinear ordinary differential equations: this method allows us to find many redundant solutions.

We affirm that Salas, Gomez and Herana\'ndez in \cite{Salas} made the errors that were discussed in works \cite{Kudryashov08, Kudryashov09, Kudryashov09a, Kudryashov09b, Kudryashov09c, Kudryashov09d, Kudryashov10, Parkes, Parkes10}.

We are grateful to one of referee for the useful remarks and his careful consideration.


\begin{thebibliography}{99}


\bibitem{Salas} Alvaro H. Salas S., Cesar A. Gomez S, Jairo Ernesto Castillo Hernan'dez New abudant solutions for the Burgers equation, Computers and Mathematics with Applications, 58 (2009) P. 514-520

\bibitem{Burgers}  Burgers J.M., A mathematical model illustrating the theory of turbulence, Advances in Applied Mechanics, 1 (1948) 171-199.


\bibitem{Rosenblatt68}   Rosenblatt M., Remark on the Burgers equation,
 Phys. Fluids. 9 (1966) 1247-1248.

\bibitem{Benton66}   Benton E.R.,  Some New Exact, Viscous, Nonsteady Solutions of Burgers' Equation,
 J. Math. Phys.  9 (1968) 1129-1136.

 \bibitem{Bateman} Bateman H. Some recent researches on the motion of fluids. Monthly
Weather Review. 1915;43:163--170.


\bibitem{Hopf51}  Hopf E., The partial differential equation $u_t+u\,u_x=u_{xx}$,
Communs. Pure Appl. Math, 3 (1950) 201-230.

\bibitem{Cole50}  Cole J.D., On a quasi-linear parabolic equation occuring in aerodynamics,
 Quart. Appl. Math. 9 (1950) 225-236.

\bibitem{Whitham}  Whitham  G B.,  Linear and Nonlinear Waves,  New York:
                    Wiley-Interscience, 1974

\bibitem{Kudryashov} Kudryashov N.A. Analitical theory of
nonlinear differential equations, Moskow - Igevsk, Institute of
computer investigations, 2004, (in Russian)

\bibitem{Glaisher}   Glaisher J.W.L., On Riccati's Equation and Its Transformations, and on Some Definite Integrals Which Satisfy Them, Phil. Trans. R. Soc. Lond, 172 (1881) 759-828.

\bibitem{Reid} Reid, William T. Riccati Differential Equations, Academic Press, New York, 1972.

\bibitem{Davis} Davis, Harold T. Introduction to Non-Linear Differential and Integral Equations,
Dover, New York, 1962.

\bibitem{Kamke}  Kamke E., Handbook on Ordinary Differential Equations [in German], Chelsea Publ.,1974

\bibitem{Polyanin03}  Polyanin A.D., Zaittsev V.F. ,  Handbook of Exact Solutions for Ordinary Differential Equations, Chapman and Hall/CRC Press, 2003, 689 - 733

\bibitem{Polyanin07}  Polyanin A.D. and  Manzhirov A.V., Handbook of Mathematics  for Engineers and Scientists, Chapman and Hall/CRC Press, 2007, 518 - 522

\bibitem{Kudryashov_2005a}  Kudryashov N.A., Simplest equation
method to look for exact solutions of nonlinear differential
equations, Chaos, Solitons and Fractals, 2005;24:1217 - 1231

\bibitem{Kudryashov_2005b}  Kudryashov N.A.,
Exact solitary waves of the Fisher equation, Physics Letters A., 2005;342:
99 - 106,

\bibitem{Kudryashov08}  Kudryashov N.A.,  Loguinova N.B., Extended simplest equation method for nonlinear differential equations , Applied Mathematics and Computation. 205 (2008) 396 - 402

\bibitem{Kudryashov09} Kudryashov N.A., Comment on: "A novel approach for solving the Fisher equation using Exp-function method", Physics Letters A, 373 (2009) 1196 - 1197


\bibitem{Kudryashov09a}  Kudryashov N.A., Loguinova N.B.,  Be careful with the Exp-function method,  Communications in Nonlinear Science and Numerical Simulation. 14 (2009),  1881-1890.

\bibitem{Kudryashov09b} Kudryashov N.A. , On "new travelling wave solutions" of the KdV and
the KdV - Burgers equations, Commun Nonlinear Sci Numer Simulat, 14
(2009), 1891-1900

\bibitem{Kudryashov09c}   Kudryashov N.A., Seven common errors in finding exact solutions of nonlinear differential equations, Communications in Nonlinear Science and Numerical Simulation. 14 (2009), 3507-3529

\bibitem{Kudryashov09d}	 Kudryashov N.A., Soukharev M.B., Popular ansatz methods and solitary wave solutions of the Kuramoto-Sivashinsky equation, Regular and Chaotic Dynamics, 14 (2009), 407-419

\bibitem{Kudryashov10}   Kudryashov N.A., Soukharev M.B. Comment on: multi soliton solution, rational solution of the Boussinesq - Burgers equation, Communications in Nonlinear Science and Numerical Simulation. (2009), doi:10.1016/j.cnsns.2009.07.016

\bibitem{Parkes} Parkes E.J., A note on travelling - wave solutions to Lax's seventh - order KdV equation, Appl Math Computation, 215 (2009), 864 - 865


\bibitem{Parkes10} Parkes E.J., A note on travelling - wave solutions to the Ostrovsky equation, Communication in Nonlinear Science and Numerical Simulation, (2009),








\end{thebibliography}
\end{document}